\documentclass[12pt]{article}

\begin{document}

\begin{titlepage}

\begin{flushright}
\end{flushright}
\vskip 2.5cm

\begin{center}
{\Large \bf Radiatively Induced Lorentz and Gauge Symmetry\linebreak
Violation in Electrodynamics with Varying $\alpha$}
\end{center}

\vspace{1ex}

\begin{center}
{\large Alejandro Ferrero\footnote{{\tt ferrero@physics.sc.edu}}
and Brett Altschul\footnote{{\tt baltschu@physics.sc.edu}}}

\vspace{5mm}
{\sl Department of Physics and Astronomy} \\
{\sl University of South Carolina} \\
{\sl Columbia, SC 29208} \\
\end{center}

\vspace{2.5ex}

\medskip

\centerline {\bf Abstract}

\bigskip

A time-varying fine structure constant $\alpha(t)$ could give rise to Lorentz-
and $CPT$-violating changes to the vacuum polarization, which would affect photon
propagation. Such changes to the effective action can
violate gauge invariance, but they are otherwise permitted. However, in the minimal
theory of varying $\alpha$, no such terms are generated at lowest order.
At second order, vacuum polarization can generate an instability---a
Lorentz-violating analogue of a negative photon mass squared
$-m_{\gamma}^{2}\propto\alpha
\left(\frac{\dot\alpha}{2\alpha}\right)^{2}\log(\Lambda^{2})$, where
$\Lambda$ is the cutoff for the low-energy effective theory.

\bigskip

\end{titlepage}

\newpage

Two exotic forms of physics beyond the standard model that have recently gotten a lot
of attention are Lorentz symmetry violations~\cite{ref-reviews} and time-dependent
fundamental constants. If these two phenomena exist, they are likely to be
closely related. For example, if the fine
structure constant $\alpha=\frac{e^{2}}{4\pi}$ is actually a function of time
$\alpha(t)$, there is naturally a preferred spacetime direction,
$\partial_{\mu}\alpha$, which violates boost invariance.
We shall investigate this potential connection, which may
also be related to violations of electrodynamical gauge invariance.

The most common theory with varying constants that has been studied is one with a
time-dependent $\alpha$. Other possibilities have included changes in the
quantum chromodynamics scale $\Lambda_{QCD}$ and therefore the
electron-proton mass ratio. However, these possibilities are usually approached
phenomenalistically, without reference to a full underlying quantum field theory.
Lorentz violation has been treated somewhat differently in recent years. Although
there is a long history of searches for deviations from special relativity---also
frequently handled in purely phenomenalistic fashion---the standard approach now is to
use effective field theory. The effective field theory containing all possible local,
Lorentz-violating operators built from standard model fields is called the standard
model extension (SME)~\cite{ref-kost2}. The SME Lagrange density contains new
operators, involving
tensor objects constructed from quantum fields, contracted with constant background
tensors. An example of such an operator is $-a^{\mu}\bar{\psi}
\gamma_{\mu}\psi$.
For parameterizing the results of
experimental tests, a restricted subset of the theory, the minimal SME, is
typically used. The minimal
SME contains only operators that are gauge invariant and power counting
renormalizable. 

Lorentz violation and varying constants have both been tightly constrained
experimentally, and many of the
most stringent tests are in quantum electrodynamics (QED). In most cases, it is
conventional to consider only the leading order effects of Lorentz
violation or a varying $\alpha$; since $\dot{\alpha}$ must be small,
${\cal O}(\dot{\alpha}^{2})$ terms may be of negligible importance. 
In this paper, we shall begin by following this convention; however, the higher order
terms are interesting when they can produce qualitatively different effects than
those possible at leading order. For this reason, after discussing
${\cal O}(\dot{\alpha})$ radiative corrections, we shall consider additional effects
that are ${\cal O}(\dot{\alpha}^{2})$ and ${\cal O}(\ddot{\alpha})$.

Measurements of $\frac{\dot{\alpha}}{\alpha}$ may be made in a number of different
ways: with pairs of precision spectroscopy experiments done years
apart~\cite{ref-fischer},
by determination of the production rates for certain isotopes in natural
reactors~\cite{ref-gould}, and by observing spectra from cosmologically distant
sources~\cite{ref-tzanavaris}.
The resulting bounds are typically at the $\frac{\dot{\alpha}}{\alpha}<10^{-14}$
yr$^{-1}$ level for measurements of the present rate of change and a comparable
$\frac{\Delta\alpha}{\alpha}<10^{-5}$ level over cosmological time scales.

Observations of very old photons can be an excellent way to probe exotic physics.
If $\alpha$ was different at the time of their emission, these photons could
reveal different characteristic atomic spectra than are seen today. The photons'
extremely long travel times could magnify the effects of a very small $\dot{\alpha}$.
If there are any changes in the propagation characteristics of the photons, this
could also be magnified by the extremely long line of sight to cosmological sources.
This is how many of the best constraints on Lorentz-violating effects have been set. For example, many SME operators give rise to a polarization dependence in the
phase speed of photons; this in turn leads to birefringence and a change in the
polarization of radiation as it propagates---an effect which has not been
seen~\cite{ref-kost22}. One can also look for energy dependence in photon arrival
times, indicative of a nontrivial photon dispersion relation~\cite{ref-galaverni},
deflection of photon trajectories~\cite{ref-kobychev}, or phase
differences~\cite{ref-altschul8}. All these can be used to probe exotic new physics
possibilities, and the long photon propagation times give tiny effects a long time to
build up, leading to many of the best bounds on various novel effects.

It is therefore worthwhile to ask whether a changing $\alpha$ would leave an imprint
on pure photon propagation. The problem is theoretically interesting, and it could
potentially lead to new ways of constraining $\dot{\alpha}$. Obviously, for a
time-dependent $\alpha$ to affect photons' propagation, the photons must interact
with some charged matter. However, it is well known that, even in vacuum, the
electromagnetic field is constantly interacting with virtual
particle-antiparticle pairs. The polarizability of the QED vacuum could
give rise to $\dot{\alpha}$-dependent effects in the pure photon sector.
And since a theory with varying $\alpha$ supports a preferred direction
$\partial_{\mu}\alpha$, it is natural to wonder whether the pure photon sector will
exhibit timelike Lorentz violation as a result of $\dot{\alpha}$.

Lorentz symmetry is closely related to $CPT$ symmetry. Breaking of $CPT$ symmetry
requires either a breaking of Lorentz invariance or something even more
exotic~\cite{ref-greenberg}.
The phenomenon of time-varying fundamental constants is odd under time reversal, even
under parity and charge conjugation---hence odd under $CPT$. The behavior under $T$
is obvious, and the behavior under $P$ is dictated by isotropy.
One might wonder if a different behavior under $C$ is possible, but to have $C$-odd,
time-dependent charges would violate the equality of particle and antiparticle
charge.

Gauge invariance is another crucial property of electrodynamics, and one which has
also been subjected to stringent experimental tests. Ordinarily, gauge symmetry is
responsible for ensuring charge conservation, but charge is not conserved if
$\dot{\alpha}\neq0$. So it would not be unexpected that radiative corrections in the
photon sector might break gauge symmetry. The most common form of gauge symmetry
breaking to be considered is a photon mass, which leads to a screening of
electromagnetic fields. Measurements of the solar wind and the persistence of the
sun's magnetic field out to Pluto's orbit give bounds at the
$m_{\gamma}<10^{-27}$ GeV level~\cite{ref-ryutov}; less secure limits based on
galaxy-scale fields are many orders of magnitude better~\cite{ref-chibisov}.
Discrete symmetries rule out the generation
of a conventional photon mass (which would be even under $C$, $P$, and $T$)
by ${\cal O}(\dot{\alpha})$ effects, but qualitatively similar Lorentz- and
gauge-symmetry-violating effects might be possible. A simple
screening of static sources would not arise, since such screening is time
reversal invariant, but changes in the magnitude of the electromagnetic field with
time are not ruled out.

Moreover, any Lorentz-violating radiative correction that is linear in $\dot{\alpha}$
must break gauge symmetry or otherwise lie outside the minimal SME.
The minimal SME contains no operator with the same discrete symmetries as
$\dot{\alpha}$. There are operators [$b_{j}$, $(k_{AF})_{j}$, $g_{jk0}$,
and $g_{j0k}$] that are $C$-even, $P$-even, and $T$-odd. However, the behavior of
Lorentz-violating effects under parity can be further subcategorized.  The parity operator is defined as inverting all three spatial coordinates,
$\vec{x}\rightarrow-\vec{x}$, but $P$ may be broken down into the product of
three separate reflections, $P=R_{1}R_{2}R_{3}$, where $R_{j}$ takes $x_{j}
\rightarrow-x_{j}$ and leaves the other two coordinates unchanged. A changing
$\alpha$ is an isotropic effect, and it must behave the same way under all three
$R_{j}$ reflections. However, the $b_{j}$, $(k_{AF})_{j}$, $g_{jk0}$,
and $g_{j0k}$ coefficients are actually odd under two
reflections and even under the third; the overall behavior under parity is the same,
but the different symmetries still prevent $\dot{\alpha}$ from generating
any of these operators through radiation corrections.

We may actually hope that gauge symmetry breaking might lead to an enhancement of
some $\dot{\alpha}$-dependent radiative corrections. New contributions to the
self-energy tensor $\Pi^{\mu\nu}$ that do not satisfy the Ward identity,
$p_{\mu}\Pi^{\mu\nu}(p)=0$ (which ensures that radiative corrections do not generate
a quadratically divergent photon mass), might naively be expected to
exhibit power law divergences. At lowest order, we would expect the divergence to
be ${\cal O}(\dot{\alpha}\Lambda)$, where $\Lambda$ is the cutoff
for the theory. A radiative correction like this could have profound experimental
implications; if the appropriate form of gauge symmetry violation could be
constrained at a scale comparable to the best bounds on $m_{\gamma}$,
the corresponding bound on $\frac{\dot{\alpha}}{\alpha}$ would be at the
$\alpha^{-1}\left(\frac{m_{\gamma}}{\Lambda}\right)m_{\gamma}$ level, with the
factor in parentheses providing a huge boost in sensitivity.
Unfortunately however, ${\cal O}(\dot{\alpha}\Lambda)$ divergences
are actually forbidden; we shall see that the radiative corrections
must depend on a positive power of $p$, which means that a linear divergence is
ruled out by power counting. But for ${\cal O}(\dot{\alpha}^{2})$ corrections the
situation is different, and radiative corrections that violate gauge invariance may
indeed be proportional to a (logarithmically) divergent function of the cutoff.

The form $\dot{\alpha}$-dependent effects might take may be model dependent, and
even if the value of $\alpha$ has changed significantly over the lifetime of the
universe, it is not really clear what the course of the time dependence might
have been. For definiteness, we shall assume a particularly simple function
$\alpha(t)$ for our discussion of ${\cal O}(\dot{\alpha})$ effects.
We shall assume that the change is smooth and uniform in time, even on very short time scales, less than the lifetime of a virtual electron-positron pair,
$\tau\sim(2m)^{-1}$.

If the electromagnetic coupling is changing on very short time scales, then we can
envision a scenario in which electromagnetic fields might grow or decay with time.
An incoming photon produces a virtual electron-positron pair, which annihilate at a
slightly later time, producing a outgoing photon. During the lifetime of the virtual
pair, the coupling constant changes slightly, and so the final field may be weaker or
stronger than the initial field was. There will, however, be cancellation between
this process and one in which the outgoing photon is produced simultaneously with the
electron and positron (which are annihilated at a later time, along with the
incoming photon).
In the second process, the change in $\alpha$ between the photon absorption and
emission has the opposite sign relative to the first.
However, the cancellation between the two should be incomplete, since the second
process has a larger energy defect in the intermediate state.

The standard vacuum polarization Feynman diagram encompasses both virtual
elec\-tron-positron processes. Normally, the Feynman rules are derived from the
action using functional derivatives. It is not automatically clear, however, whether
a theory with varying constants should be derived from a Lagrangian. 
In fact, there are ambiguities in the Lagrangian formalism.
The standard electromagnetic Lagrange density may be written as
\begin{equation}
{\cal L}_{EM}=-\frac{1}{4e^{2n}}F^{\mu\nu}F_{\mu\nu}+e^{1-n}j^{\mu}A_{\mu}
\end{equation}
for any $n$. When $e$ is constant, the value of $n$ is irrelevant. However, when the
coupling varies with time, only the Lagrangian
with $n=1$ is invariant under the usual
$U(1)$ gauge transformation. The equation of motion becomes
\begin{equation}
\label{eq-ofmotion}
\frac{1}{e^{2n}}\partial_{\mu}F^{\mu\nu}-\frac{4\pi
n}{e^{2n+2}}(\partial_{\mu}\alpha)F^{\mu\nu}=
e^{1-n}j^{\nu}.
\end{equation}
This includes a Lorentz-violating term with an undetermined coefficient. Theories
of this sort are discussed in~\cite{ref-kost20}.
The Lorentz-violating term in (\ref{eq-ofmotion})
mirrors the effects of a gauge-noninvariant $\vec{A}\cdot\vec{E}$ term in ${\cal L}$.
However, since the undetermined dependent coefficient can be
set to zero (using the very the common choice of $n=0$), it is possible to
have a theory with varying $\alpha(t)$ and no Lorentz violation of this
sort.
Moreover, the differences between the theories with different values of $n$ are not
generally apparent in experiments that
simply compare the instantaneous value of $\alpha$
at two different times.

Rather than dealing with the Lagrangian, we
shall simply take the theory to be defined by its Feynman rules. This allows us to
consider the minimal modification to QED that accounts for a
time-varying $\alpha$. However, this approach is essentially equivalent to using
the particular electromagnetic Langragian with $n=1$.

Because the coupling constant, and thus the Feynman rules,
will be time dependent, we shall set up the rules
in configuration space. The key change to the QED Feynman rules is in the vertex; a
vertex at position $x$ corresponds to a factor $-\int\! d^{4}x\,ie\left(1+
\frac{\dot{\alpha}}{2\alpha}x^{0}\right)\gamma^{\mu}$. Any theory with varying
$\alpha$ essentially must include this modified vertex. We shall take the factors
corresponding to other elements of Feynman graphs to be unchanged. A fully
consistent theory might include additional modifications to the Feynman rules,
but we want to consider only those effects which are necessary consequences of having
a time-dependent $\alpha$.
The remaining rules
that will be needed for the calculation of the modified photon self-energy are the
factors of $e^{-iq\cdot x}\epsilon_{\mu}$ ($e^{iq\cdot x}\epsilon^{*}_{\mu}$) for
incoming (outgoing) photon lines and the fermion propagator
\begin{equation}
S_{F}(x-y)=\int\!\frac{d^{4}p}{(2\pi)^{4}}\frac{i(\not\! p\,+m)}{p^{2}-m^{2}
-i\varepsilon}e^{-ip\cdot(x-y)},
\end{equation}
with a $-1$ for a closed fermion loop.

Novel effects in the pure photon sector would be generated by the vacuum polarization
diagram.
Excising the external photon lines and the corresponding polarization vectors, the
amplitude for this diagram is
\begin{eqnarray}
i{\cal M}^{\mu\nu}(q,q') & = & \int\! d^{4}x
\int\! d^{4}y\int\!\frac{d^{4}k_{1}}{(2\pi)^{4}}
\int\!\frac{d^{4}k_{2}}{(2\pi)^{4}}(-ie)^{2}\left(1+\frac{\dot{\alpha}}{2\alpha}x^{0}
\right)\left(1+\frac{\dot{\alpha}}{2\alpha}y^{0}\right) \\
& & \times (-1)\,{\rm tr}\left[\gamma^{\mu}
\frac{i(\not\! k_{1}\,+m)}{k_{1}^{2}-m^{2}}
\gamma^{\nu}\frac{i(\not\! k_{2}\,+m)}{k_{2}^{2}-m^{2}}\right]
e^{-i(k_{1}-k_{2})\cdot(x-y)}e^{-iq\cdot x}e^{iq'\cdot y}. \nonumber
\end{eqnarray}
The integrals may be rewritten in
terms of $k=k_{1}$ and $p=k_{2}-k_{1}$. Symmetries dictate that, when we calculate
only the fermion-antifermion loop, the $x^{0}$ and $y^{0}$ contributions are
equal apart from surface terms.
For the ${\cal O}(\dot{\alpha})$ part of the self-energy, we therefore have
\begin{eqnarray}
i{\cal M}^{\mu\nu}_{\dot{\alpha}}(q,q') & = &
e^{2}\left(\frac{\dot{\alpha}}{\alpha}\right)
\int\! d^4{x}\int\! d^{4}y\, y^{0}\int\!\frac{d^{4}p}{(2\pi)^{4}}
e^{ip\cdot(x-y)}e^{-iq\cdot x}e^{iq'\cdot y} \\
& & \times\int\!\frac{d^{4}k}{(2\pi)^{4}}\,{\rm tr}
\left[\gamma^{\mu}\frac{i(\not\! k\,+m)}{k^{2}-m^{2}}
\gamma^{\nu}\frac{i(\not\! k\,+\not\! p\,+m)}{(k+p)^{2}-m^{2}}\right] \nonumber\\
\label{eq-FOselfE}
& = & \left(\frac{\dot{\alpha}}{\alpha}\right)
\int\! d^4{x}\int\! d^{4}y\, y^{0}\int\!\frac{d^{4}p}{(2\pi)^{4}}
e^{ip\cdot(x-y)}e^{-iq\cdot x}e^{iq'\cdot y}\left[i\Pi^{\mu\nu}_{2}(p)\right],
\end{eqnarray}
where $\Pi^{\mu\nu}_{2}(p)$ is the usual one-loop photon self-energy.

The $y$-integration may be eliminated via $\int\! d^{4}y\,y^{0}
e^{-ip\cdot y}e^{iq'\cdot y}=i(2\pi)^{4}\delta^{3}(\vec{p}-\vec{q}\,')\delta'(p^{0}-
q'^{0})$. This reduces the self-energy to
\begin{eqnarray}
i{\cal M}^{\mu\nu}_{\dot{\alpha}}(q,q') & = &
\left(\frac{\dot{\alpha}}{\alpha}\right)\int\! d^{4}x
\int\! d^{4}p\,ie^{-i(\vec{p}-\vec{q}\,)\cdot\vec{x}}e^{i(p^{0}-q^{0})x^{0}}
\delta^{3}(\vec{p}-\vec{q}\,')\delta'(p^{0}-q'^{0})\left[i\Pi^{\mu\nu}_{2}(p)\right]
\nonumber\\
& = & \left(\frac{\dot{\alpha}}{\alpha}\right)
(2\pi)^{3}\delta^{3}(\vec{q}-\vec{q}\,')\int\! dx^{0}
\int\! dp^{0}\, ie^{i(p^{0}-q^{0})x^{0}}\delta'(p^{0}-q'^{0})
\left[i\Pi^{\mu\nu}_{2}(p^{0},\vec{q}\,)\right] \\
\label{eq-selfe}
& = & \left(\frac{\dot{\alpha}}{\alpha}\right)
(2\pi)^{3}\delta^{3}(\vec{q}-\vec{q}\,')\int\! dx^{0}\Biggl\{x^{0}e^{-i(q^{0}-q'^{0})
x^{0}}\left[i\Pi^{\mu\nu}_{2}(q^{0},\vec{q}\,)\right] \\
& & +\,e^{-i(q^{0}-q'^{0})x^{0}} \left[\frac{\partial\Pi^{\mu\nu}_{2}(q)}
{\partial q^{0}}\right]\Biggr\} \nonumber
\end{eqnarray}
It is not possible to factor out an energy-conserving $\delta$-function, because
the theory is not time translation invariant. However,
the first term in French brackets in (\ref{eq-selfe}) has a straightforward
interpretation. Even though it appears to violate energy conservation, it is
actually a natural extension of the usual photon self-energy.
Combined with the $\dot{\alpha}$-independent
contribution, it gives a contribution
to the effective Lagrangian
$\Delta{\cal L}
\supset\frac{1}{2}\left(1+\frac{\dot{\alpha}}{\alpha}x^{0}\right)A_{\mu}(x)
[\Pi_{2}^{\mu\nu}(i\partial)]A_{\nu}(x)$. This just represents the usual
quantum correction, but with the full, time-dependent $\alpha(t)$. A crucial feature
of $\Pi_{2}^{\mu\nu}$ is that it obeys the Ward identity.
It is this property [and the fact that the self-energy is regular at
$p^{2}=0$] that ensures photons are massless. The first term in (\ref{eq-selfe})
also obeys the Ward identity.

If we had retained the two external photon propagators in the
preceding calculation and determined the amplitude for a photon
to propagate from $z_{1}$ to
$z_{2}$, splitting into an electron-positron pair one time along the way, the result
would be
just the usual expression, but with the coupling constant $\alpha$ evaluated at
the average time $(z_{1}^{0}+z_{2}^{0})/2$.

The fact that the effective Lagrangian depends explicitly on time suggests an
interesting possibility. The effective Hamiltonian for the electromagnetic field also
appears to be time dependent, which could lead to energy nonconservation. However,
this apparent effect is actually unphysical. The time dependence of the effective
Hamiltonian density ${\cal H}+\Delta{\cal H}$ arises from the time dependence of
$\Delta{\cal L}$, which is exactly canceled by a
time-dependent field strength renormalization constant $Z(t)$; the
instantaneous value of $Z(t)$ sets the scale on which field strengths are measured.
That such a cancellation must occur is actually clear from the fact that the
momentum density of the field appears to vary in time in exactly the same way that
the Hamiltonian density does. However, the theory is invariant under spatial translations, and so momentum must be conserved; any apparent time variation is
canceled by $Z(t)$.

The second term on the right-hand side of (\ref{eq-selfe}) is superficially Lorentz-,
$CPT$-, and gauge-symmetry-vi\-o\-la\-ting, but it
is actually a total derivative. It gives rise to terms such as
$A^{\mu}\partial^{0}A_{\mu}=\frac{1}{2}\partial^{0}A^{2}$ and
$A^{0}(\partial_{\mu}A^{\mu})+A^{\mu}(\partial_{\mu}A^{0})=
\partial_{\mu}(A^{0}A^{\mu})$,
which certainly cannot generate a novel time dependence for electromagnetic fields,
nor any other new effects. In fact, had we performed the various integrations in a
different order, this term need not have appeared at all.

There are several reasons why the radiative corrections take these kinds of
forms. In the physical picture outlined earlier, there was a partial cancellation
between processes in which the incoming field was absorbed before the outgoing one
was emitted and processes with the time ordering inverted. If the incoming field carried no energy, the intermediate states in the two processes would be equally
off shell, and the cancellation would be exact;
therefore, any nonvanishing contribution
must be proportional to the photon energy $p^{0}$. For a radiative
correction to produce the notional time dependence
(with a growing or decaying field), it must lead to an equation of
motion that may be written schematically as $\ddot{A}-\xi\dot{\alpha}
(\partial)^{2n+1} A=0$. The $\dot{\alpha}$ term must have an odd number of
derivatives, so that changing the sign of $\dot{\alpha}$ will reverse the decay or
enhancement effect. Alternatively, having the radiative correction depend on an odd
power of the momentum is the only way to preserve the  $C$-even, $P$-even, $T$-odd
behavior characteristic of ${\cal O}(\dot{\alpha})$ effects.

There are terms
in the SME with odd powers of $p$ that are not total derivatives---for example,
the purely timelike Chern-Simons term $\frac{1}{2}k_{AF}^{0}\vec{A}\cdot\vec{B}$.
However, this
term relies on the presence of the Levi-Civita $\epsilon$-tensor in the
definition of the magnetic field and is consequently odd under parity; it affects
right- and left-circularly polarized photons in opposite fashions.
Moreover, the Chern-Simons term is gauge
invariant (up to a total divergence). However, a gauge symmetry breaking term such as
$A^{0}(\partial^{\mu}A_{\mu})$ could make equally
real contributions to photon behavior, if it appeared singly (and not in
combination as a total derivative). In that case, the term would duplicate the
effects of the Lorentz-violating term in (\ref{eq-ofmotion}).

Since one of our major
results is that the variation of $\alpha$ does not give rise to a
Lorentz-violating two-photon operator at lowest order,
it is natural to wonder what was wrong with the scenario we described earlier,
with the change in the coupling constant leading to temporal
enhancement or suppression of electromagnetic fields. In that scenario, we
would expect the amplitude of the field to track the value of $\alpha(t)$;
as a photon is continually being absorbed and re-emitted by virtual
electron-positron pairs, the field strength should increase (or decrease)
as the charge does. However, this is
problematic when considered
in the context of quantum mechanics. When a single photon is propagating,
a continuous change in the magnitude of the field strength is not
possible; such a change would violate the quantization of energy and hence
the uncertainty principle. Rather than the field strength, what tracks the
changing value of $\alpha$ is the
amount of mixing between photon and fermion-antifermion states.
This argument also suggests that there
ought to be no Lorentz-violating radiative corrections
at ${\cal O}(\dot{\alpha})$ even when diagrams with more loops are considered.

We shall now delve into the question of what happens at higher orders, either
${\cal O}(\dot{\alpha}^{2})$ or ${\cal O}(\ddot{\alpha})$. In a systematic
expansion of the photon self-energy, these two orders should be considered
simultaneously. We shall take the time-dependent electric charge $e(t)$ as simply
proportional to the square root of $\alpha(t)$:
\begin{equation}
\frac{e(t)}{\sqrt{4\pi\alpha_{0}}}=1+\left(\frac{\dot{\alpha}}{2\alpha_{0}}\right)t
+\frac{1}{2}\left[\left(\frac{\ddot{\alpha}}{2\alpha_{0}}\right)-
\left(\frac{\dot\alpha}{2\alpha_{0}}\right)^{2}\right]t^{2},
\end{equation}
where $\alpha_{0}$ is the fine structure constant at the reference time $t=0$.
However, calculations that include these higher order terms do not enjoy the model
independence of the ${\cal O}(\dot{\alpha})$ results; it would be entirely natural to
have other changes to the Feynman rules at this order.
It is also important to note that the ${\cal O}(\dot{\alpha}^{2})$
radiative corrections may be smaller
than corrections that are only ${\cal O}(\dot{\alpha})$, but which arise from
diagrams with more than one loop---although the terms at the different orders can
always be distinguished by their behavior under $T$.

With the higher order corrections, we encounter a new subtlety. To determine the
one-loop effective action in this theory,
it is not generally sufficient to evaluate an isolated fermion-antifermion loop,
amputated of external legs. This complication arises because of the explicit time
dependence in the Feynman rules.
The factors of $x^{0}$ lead to derivatives of $\delta$-functions;
these produce $\partial/\partial q_{0}$ derivative operators which act not
only on $\Pi^{\mu\nu}(q)$, but also on the external propagators in the
photon self-energy diagram. This subtlety does not become a problem at
${\cal O}(\dot{\alpha})$, where the full results may be inferred from just
the fermion-antifermion part of the self-energy. However, at
higher orders, we must calculate the full two-point correlation function
$\langle A^{\mu}(z_{1})A^{\nu}(z_{2})\rangle$.

The calculation of this two-point function is outlined in the Appendix. The
result, including all terms up to ${\cal O}(\dot{\alpha}^{2})$ and
${\cal O}(\ddot{\alpha})$ is
\begin{eqnarray}
\langle A^{\mu}(z_{2})A^{\nu}(z_{1})\rangle
& = & \frac{\bar{\alpha}}{\alpha_{0}}\int\!\frac{d^{4}q}{(2\pi)^{4}}
\frac{-i}{q^{2}}\frac{-i}{q^{2}}e^{-iq\cdot(z_{2}-z_{1})}\left[i\Pi_{2}^{\mu\nu}(q)
\right] \nonumber\\
& & +\,\frac{1}{8}\left(\frac{\dot{\alpha}}{\alpha_{0}}\right)^{2}
\int\!\frac{d^{4}q}
{(2\pi)^{4}}\frac{-i}{q^{2}}\frac{-i}{q^2}e^{-iq\cdot(z_{2}-z_{1})}
\left[i\frac{\partial^{2}}{\partial q_{0}^{2}}\Pi_{2}^{\mu\nu}(q)\right] \nonumber\\
& & +\,\frac{1}{2}\left(\frac{\ddot{\alpha}}{\alpha_{0}}\right)
\int\!\frac{d^{4}q}{(2\pi)^{4}}
\frac{-i}{q^{2}}\frac{-i}{q^{2}}e^{-iq\cdot(z_{2}-z_{1})}\left\{iq_{0}
\frac{\partial}{\partial q_{0}}
\left[\frac{\Pi_{2}^{\mu\nu}(q)}{q^{2}}\right]\right\},
\label{eq-completeselfE}
\end{eqnarray}
where $\bar{\alpha}$ is a suitably averaged value of the coupling between the
times $z_{1}^{0}$ and $z_{2}^{0}$:
\begin{equation}
\frac{\bar{\alpha}}{\alpha_{0}}
=1+\frac{\dot{\alpha}}{2\alpha_{0}}(z_{1}^{0}+z_{2}^{0})+\frac{\ddot{\alpha}}
{4\alpha_{0}}\left[\left(z_{1}^{0}\right)^{2}+\left(z_{2}^{0}\right)^{2}\right].
\end{equation}

Equation (\ref{eq-completeselfE})
factorizes the two-point function into approximately the form usually seen.
It includes the usual self-energy, modified by an average of the
time-dependent coupling constant, and also new, Lorentz-violating self-energy
terms. We can almost just read off the effective potential, as we would
conventionally.
However, when multiple fermion-antifermion loops are inserted into the
photon propagator, the loops do not quite produce an exactly resummable geometric
series. The existence of the $\partial/\partial q_{0}$ derivatives acting
on the several parts of the diagram is responsible for this. However, the
terms do approximately resum; the errors associated with the
resummation are ${\cal O}(\alpha^{2})$, equivalent to other multiple-loop
corrections that have already been neglected.
So at the order under consideration, the
${\cal O}(\dot{\alpha}^{2})$ and ${\cal O}(\ddot{\alpha})$ terms can be treated
like any other Lorentz-violating contribution to the effective Lagrangian.

The ${\cal O}(\dot{\alpha}^{2})$ part of the
self-energy is
\begin{equation}
\Pi_{m_{\gamma}}^{\mu\nu}(q)
=\frac{1}{8}\left(\frac{\dot{\alpha}}{\alpha_{0}}\right)^{2}
\left[\frac{\partial^{2}}{\partial q_{0}^{2}}\Pi_{2}^{\mu\nu}(q)\right],
\end{equation}
which has the general structure of a photon mass term.
The discrete symmetries that prevented the generation of a mass term at first order
do not come into play here. However, this mass-like
contribution differs from the usual
Proca mass term~\cite{ref-proca} and clearly breaks Lorentz boost symmetry.

The Lorentz structure of $\Pi^{\mu\nu}_{2}(q)$ is
$\Pi^{\mu\nu}_{2}(q)=(q^{2}g^{\mu\nu}-q^{\mu}q^{\nu})\Pi_{2}(q^{2})$. The scalar
factor is $\Pi_{2}(q^{2})=\Pi_{2}(0)+\frac{2\alpha}{\pi}\int_{0}^{1}\!dz\,z(1-z)\log
\left[1-z(1-z)q^{2}/m^{2}\right]$. The constant term in $\Pi_{2}(q^{2})$ is formally
divergent and in any case should be much larger than the momentum-dependent term when
$q^{2}\ll m^{2}$. So the dominant part of the radiative correction is captured in the
approximation $\Pi_{2}(q)\approx\Pi_{2}(0)$. The second derivative then becomes
$\frac{\partial^{2}}{\partial q_{0}^{2}}\Pi_{2}^{\mu\nu}(q)\approx
2(g^{\mu\nu}-g^{0\mu}g^{0\nu})\Pi_{2}(0)$, and the ${\cal O}(\dot{\alpha}^{2})$
contribution to the effective Lagrange density is
\begin{equation}
\Delta{\cal L}\supset-\frac{1}{8}
\left(\frac{\dot\alpha}{\alpha_{0}}\right)^{2}\Pi_{2}(0)\vec{A}\,^{2}.
\end{equation}

Of course, this expression is problematic, because $\Pi_{2}(0)$ is formally
infinite. The divergence itself is not surprising, since the term involved is not
gauge invariant; however, we must understand what the physical interpretation of this
divergence could be.
The mass-like
term represents a new renormalizable operator, which is consistent with the
broken gauge and Lorentz symmetries of the theory. The term will be generated by
radiative corrections even if its bare value vanishes, and naturalness suggests
that the physically measurable coefficient of the term should not be substantially smaller in magnitude than the radiative corrections it receives.
With a single species of fermion, the divergence associated with $\Pi_{2}(0)$
takes the form $\Pi_{2}(0)=-\frac{\alpha_{0}}{3\pi}\log\frac{\Lambda^{2}}{m^{2}}$,
where
$\Lambda$ is the ultraviolet cutoff. What value of $\Lambda$ is appropriate depends
on the scale at which the new physics ultimately responsible for the varying
$\alpha$ enters, but the cutoff should presumably lie somewhere between the
electroweak scale and the Planck scale. Taking $\Lambda=200$ GeV as a conservative
estimate and including all species of charged fermions gives
$\left|\Pi_{2}(0)\right|>10\alpha_{0}$.

A mass term of the form $-m_{\gamma}^{2}\vec{A}\,^{2}$ has interesting
properties~\cite{ref-gabadadze,ref-dvali}. While many theories that invoke Lorentz
violation envision it occurring primarily as a high-energy phenomenon, this form
of Lorentz violation is most important far in the infrared (just like the changing
$\alpha$ that generates it).
With this mass term, the magnetic field of a source takes the
same form as in Proca electrodynamics, while the electric field
exhibits a mixture of Proca and conventional behavior. There is an instantaneous
Coulomb's Law, while propagating waves have a massive dispersion relation. The signal
speed for the theory is infinite, but electromagnetic disturbances caused by distant
sources are slow to reach full strength.
Constraints on the photon mass based on observations of static magnetic fields would
remain valid even if the mass has this Lorentz-violating form, but constraints based
on the behavior of static electric fields would not.

Of course, the radiative correction is not actually a photon mass term, because
$m_{\gamma}^{2}$ is negative. There are also the derivatives acting on
$\Pi_{2}(q^{2})-\Pi_{2}(0)$. These give additional smaller
corrections---some of which are gauge invariant, some of which are not.
However, these terms will not change the overall sign of the mass-like term.
Because of its dependence on $\dot{\alpha}^{2}$, this term cannot represent a
genuine mass, whether the coupling strength is increasing or decreasing. Instead,
it should correspond to some kind of instability.
The existence of either a mass term or an instability at
${\cal O}(\dot{\alpha}^{2})$ would suggest the possibility of constraining $\dot{\alpha}$ indirectly, using large scale electromagnetic field measurements.
However, the best limits on the time dependence of $\alpha$ are
already many orders of magnitude better than the best current limits on
$m_{\gamma}$ and related quantities, so direct measurements cannot provide any useful
new constraints on $\dot{\alpha}$. In fact, the bounds on the variation of $\alpha$
are so tight
that it is not clear that the usual interpretation assigned to an instability is
still meaningful. An instability on a time scale longer than the lifetime of
the universe would not be directly measurable (and with a positive $m_{\gamma}^{2}$ of the same size, there would be a similar problem; the energy uncertainty of any
observed photon cannot be less than the reciprocal lifetime of the universe).
So the presence of this mass-like term is theoretically interesting but probably not
experimentally significant.

The Lorentz-violating term that arises at ${\cal O}(\ddot{\alpha})$ has a different
structure.
Again neglecting $\Pi_{2}(q^{2})-\Pi_{2}(0)$, this term is equivalent to a
self-energy contribution
\begin{equation}
\Pi_{\ddot{\alpha}}^{\mu\nu}(q)=\frac{\ddot{\alpha}}{2\alpha_{0}}\left[
-\frac{g^{\mu0}q_{0}q^{\nu}+g^{\nu0}q_{0}q^{\mu}}{q^{2}}+\frac{2q_{0}^{2}q^{\mu}
q^{\nu}}{(q^{2})^{2}}\right]\Pi_{2}(0).
\end{equation}
This term violates gauge invariance---$q_{\mu}\Pi_{\ddot{\alpha}}^{\mu\nu}(q)
\propto g^{\nu0}q_{0}-q_{0}^{2}q^{\nu}/q^{2}$.
However, it is not a photon mass term.
At $\vec{q}=0$, it vanishes; it does not endow long-wavelength excitations
with a nonzero energy. Nor, at leading order, does it affect the
dispersion relation for standard photons; contracted with a transverse
spatial polarization vector, it also gives 0. Indeed, if this term is inserted
(as part of the photon propagator) between two conserved currents, it will always
give a vanishing result. The term is not
completely trivial in a theory that incorporates charge nonconservation, but any
physical effects will be additionally suppressed by the smallness of
$\partial^{\mu}j_{\mu}$---which, in this theory, is ${\cal O}(\dot{\alpha})$. So
this term only has effects at higher order.

At ${\cal O}(\ddot{\alpha})$, the additional terms arising from derivatives of
$\Pi_{2}(q^{2})$ are all transverse and thus gauge invariant. The largest correction
of this type comes from
\begin{equation}
\left.\frac{q_{0}}{q^{2}}\frac{\partial\Pi_{2}(q^{2})}{\partial q_{0}}
\right|_{q^{2}\rightarrow0}=-\frac{2\alpha_{0}q_{0}^{2}}{15\pi m^{2}q^{2}}.
\end{equation}
On its own, this term would modify the dispersion relation in the denominator of
the photon propagator to $q^{2}+\left(\frac{\ddot{\alpha}}{\alpha_{0}}\right)
\frac{\alpha_{0}}{15\pi m^{2}}q_{0}^{2}$. This represents an
isotropic change in the propagation speed of photons and is a radiative contribution
to the SME parameter $\tilde{\kappa}_{{\rm tr}}$, with
$\Delta\tilde{\kappa}_{{\rm tr}}=\left(\frac{\ddot{\alpha}}{\alpha_{0}}\right)
\frac{\alpha_{0}}{30\pi m^{2}}$. This shows that a varying $\alpha(t)$ does generate
quantum corrections to minimal SME operators.
However, given the experimentally allowed values
of $\ddot{\alpha}$ and the size of the electron mass, this correction must be
incredibly minuscule and not observable directly.

In summary,
we have studied the radiative corrections to the photon sector caused by a time
varying fine structure constant $\alpha$. If $\dot{\alpha}\neq0$, the theory is not
invariant under Lorentz or $CPT$ symmetry, nor is charge conserved. So there is
expected to be no symmetry preventing the appearance of Lorentz-
and gauge-symmetry-violating terms in the effective action. However, in a theory that
has been minimally modified to include a varying $\alpha(t)$, the one-loop
vacuum polarization does not generate any Lorentz-violation in the photon sector
at ${\cal O}(\dot{\alpha})$. In other words, it is possible to have a varying
$\alpha$ that is not accompanied by any electromagnetic Lorentz violation at lowest
order.
At ${\cal O}(\dot{\alpha}^{2})$, a $CPT$-preserving but Lorentz-violating
photon-mass-like term is possible, and in the minimal varying $\alpha$ theory, such a
term is duly generated.

\section*{Acknowledgments}
The authors are grateful to V. A. Kosteleck\'{y} for helpful comments.

\section*{Appendix: Self-Energy at Second Order}

The amplitude for the electromagnetic field to propagate from $z_{1}$ to $z_{2}$
with one fermion-antifermion loop insertion is
\begin{eqnarray}
\label{eq-SOselfE}
\langle A^{\mu}(z_{2})A^{\nu}(z_{1})\rangle & = & \int\!\frac{d^{4}q}{(2\pi)^4}
\int\!\frac{d^{4}q'}{(2\pi)^4}\int\!\frac{d^{4}p}{(2\pi)^4}\int\! d^{4}x\int\! d^{4}y
\frac{-ig^{\mu}\,_{\alpha}}{q^{2}}\frac{-ig_{\beta}\,^{\nu}}{q'^2}  \nonumber\\
& & e^{-iq\cdot(x-z_{1})}e^{ip\cdot(y-x)}e^{-iq'\cdot(z_{2}-y)}\left[1+
\frac{\dot\alpha}{2\alpha_0}(x_{0}+y_{0})\right. \nonumber\\
& & +\left.\frac{\dot\alpha^2}{4\alpha_{0}^{2}}x_{0}y_{0}
+\frac{1}{2}\left(\frac{1}{2}\frac{\ddot\alpha}{\alpha_0}-\frac{1}{4}\frac{\dot\alpha^2}{\alpha_0^2}\right)\left(x_{0}^2+y_{0}^2\right)\right]
\left[i\Pi_{2}^{\alpha\beta}(p)\right].
\end{eqnarray}
This is correct to second order in
the variation of the coupling constant and represents
a straightforward generalization of the first-order expression
(\ref{eq-FOselfE}).
Defining
\begin{eqnarray}
M^{\mu\nu}(\xi) & = & \int\!\frac{d^{4}q}{(2\pi)^{4}}\int\!\frac{d^{4}q'}{(2\pi)^{4}}
\int\!\frac{d^{4}p}{(2\pi)^4}\int\! d^{4}x\int\! d^{4}y\,\xi \nonumber\\
& & \frac{-i}{q^2}\frac{-i}{q'^2}e^{iq\cdot z_{1}}e^{i(p-q)\cdot x}
e^{-i(p-q')\cdot y}e^{-iq'\cdot z_{2}}\left[i\Pi_{2}^{\alpha\beta}(p)\right],
\end{eqnarray}
the ${\cal O}(\dot{\alpha})$ calculation becomes nothing more than the evaluation of
$M^{\mu\nu}(x_{0}+y_{0})$.

The ${\cal O}(\dot{\alpha}^{2})$ and ${\cal O}(\ddot{\alpha})$ terms are trickier.
In order to obtain the fewest extraneous surface terms and otherwise simplify the
calculation, we can perform the necessary
integrations in the most symmetric fashion possible. We begin with
\begin{equation}
\label{a18}
M^{\mu\nu}(x_{0}y_{0}) = \int\!\frac{d^{4}q}{(2\pi)^{4}}\int\!\frac{d^{4}q'}
{(2\pi)^{4}} \int\!\frac{d^{4}p}{(2\pi)^4}\int\! d^{4}x\int\! d^{4}y\,x_{0}y_{0}
f(q_{0},q'_{0})e^{i(p-q)\cdot x}e^{-i(p-q')\cdot y}\left[i\Pi_{2}^{\mu\nu}(p)\right],
\end{equation}
where
\begin{equation}
f(q_{0},q'_{0})=\frac{-i}{q_{0}^{2}-\vec{q}\,^{2}}\frac{-i}{q_{0}'^{2}-\vec{q}\,^{2}}
e^{iq_{0}z_{1}^{0}}e^{-iq'_{0}z_{2}^{0}}e^{i\vec{q}\cdot(\vec{z}_{2}-\vec{z}_{1})}.
\end{equation}
If we perform the integral over $x$ first, it will produce a factor of $(2\pi)^{4}i\delta^{3}(\vec{q}-\vec{p}\,)\frac{1}{2}(\partial_{q_{0}}-\partial_{p_{0}})
\delta(q_{0}-p_{0})$. If the integral over $y$ is done first, the contribution would
instead be $(2\pi)^{4}i\delta^{3}(\vec{p}-\vec{q}')\frac{1}{2}(\partial_{p_{0}}-
\partial_{q'_{0}})\delta(p_{0}-q_{0}')$. Averaging the contributions from the two
orders of integration, we find
\begin{eqnarray}
M^{\mu\nu}(x_{0}y_{0}) & = & \frac{1}{4}\int\!\frac{d^{4}q}{(2\pi)^{4}}\int\!
\frac{d^{4}q'}
{(2\pi)^{4}} \left\{\int\! d^{4}y\,e^{-i(q-q')\cdot y}\left[i\,y_{0}^{2}
\Pi_{2}^{\mu\nu}(q)f(q_{0},q'_{0})\right.\right. \\
& & -\,y_{0}f(q_{0},q'_{0})\partial_{q_{0}}\Pi_{2}^{\mu\nu}(q)
+\left. y_{0}\Pi_{2}^{\mu\nu}(q)\partial_{q_{0}}f(q_{0},q'_{0})\right]
+\int\! d^{4}x\, e^{-i(q-q')\cdot x} \nonumber\\
& & \left.\left[ix_{0}^{2}\Pi_{2}^{\mu\nu}(q')f(\omega,\omega')+x_{0}f(q_{0},q'_{0})
\partial_{q'_{0}}\Pi_{2}^{\mu\nu}(q')-x_{0}\Pi_{2}^{\mu\nu}(q')\partial_{\omega'}
f(q_{0},q'_{0})\right]\right\}. \nonumber
\end{eqnarray}
This can be written as $M^{\mu\nu}(x_{0}y_{0})=\frac{1}{4}
M^{\mu\nu}(x_{0}^{2}+y_{0}^{2})+\frac{1}{4}\bar{M}^{\mu\nu}(x_{0}y_{0})$, where

%
\begin{eqnarray}
\bar{M}^{\mu\nu}(x_{0}y_{0}) & = & \int\!\frac{d^{4}q}{(2\pi)^{4}}\int\!
\frac{d^{4}q'}{(2\pi)^{4}}\int\! d^{4}y\,y_{0}
e^{-i(q_{0}-q_{0}')y_{0}}e^{i(\vec{q}-\vec{q}')\cdot\vec{y}}
\left\{\Pi_{2}^{\mu\nu}(q)\partial_{q_{0}}f(q_{0},q'_{0})\right. \nonumber\\
& & -\left.\Pi_{2}^{\mu\nu}(q')\partial_{q'_{0}}f(q_{0},q'_{0})
+f(q_{0},q'_{0})[\partial_{q'_{0}}\Pi_{2}^{\mu\nu}(q')-\partial_{q_{0}}
\Pi_{2}^{\mu\nu}(q)]\right\} \\
& = & \frac{i}{2}\int\!\frac{d^4q}{(2\pi)^4}\left\{
2f(q_{0},q_{0})\partial^{2}_{q_{0}}\Pi_{2}^{\mu\nu}(q)
-\left[\partial_{q_{0}}f(q_{0},q_{0})\right]
\left[\partial_{q_{0}}\Pi_{2}^{\mu\nu}(q)\right]\right. \nonumber\\
& & -\left.\left.\Pi_{2}^{\mu\nu}(q)\partial^{2}_{q_{0}}f(q_{0},q_{0})
+4\Pi_{2}^{\mu\nu}(q)\left[\partial_{q_{0}}\partial_{q'_{0}}f(q_{0},q'_{0})\right]
\right|_{q'_{0}=q_{0}}\right\}.
\end{eqnarray}

Similarly,
%
\begin{eqnarray}
M^{\mu\nu}(x_{0}^{2}+y_{0}^{2}) & = & 
-i\int\!\frac{d^{4}q}{(2\pi)^{4}}\left\{f(q_{0},q_{0})
\partial^{2}_{q_{0}}\Pi_{2}^{\mu\nu}(q)+\left[\partial_{q_{0}}f(q_{0},q_{0})\right]
\left[\partial_{q_{0}}\Pi_{2}^{\mu\nu}(q)\right]\right. \nonumber\\
& & +\left.\left.\Pi_{2}^{\mu\nu}(q)\partial^{2}_{q_{0}}f(q_{0},q_{0})
-2\Pi_{2}^{\mu\nu}(q)\left[\partial_{q_{0}}\partial_{q'_{0}}f(q_{0},q'_{0})\right]
\right|_{q'_{0}=q_{0}}\right\}.
\end{eqnarray}
Tabulating the derivatives of $f$ that appear, we have
\begin{eqnarray}
f(q_{0},q_{0}) & = &\frac{-i}{q^2}\frac{-i}{q^2}\,e^{-iq\cdot(z_{2}-z_{1})} \\
\partial_{q_{0}}f(q_{0},q_{0})
& = & f(q_{0},q_{0})\left[-i(z_{2}^{0}-z_{1}^{0})-\frac{4q_{0}}{q^{2}}\right] \\
\partial^{2}_{q_{0}}f(q_{0},q_{0}) & = & f(q_{0},q_{0})
\left[\left(iz_{2}^{0}-iz_{1}^{0}+\frac{4q_{0}}{q^{2}}\right)^{2}
-\frac{4}{q^{2}}+\frac{8q_{0}^{2}}{(q^{2})^{2}}\right] \\
\left.\left[\partial_{q_{0}}\partial_{q'_{0}}f(q_{0},q'_{0})\right]
\right|_{q'_{0}=q_{0}} & = & 
f(q_{0},q_{0})\!\left(-iz_{2}^{0}-\frac{2q_{0}}{q^{2}}\right)\!\!\left(iz_{1}^{0}
-\frac{2q_{0}}{q^{2}}\right).
\end{eqnarray}
Combining these gives expressions for $M^{\mu\nu}(x_{0}^{2}+y_{0}^{2})$ and
$\bar{M}^{\mu\nu}(x_{0}y_{0})$. By adding the total derivatives
$\partial^{2}_{q_{0}}[\Pi_{2}^{\mu\nu}(q)f(q_{0},q_{0})]$ and
$\partial_{q_{0}}[f(q_{0},q_{0})\partial_{q_{0}}\Pi_{2}^{\mu\nu}(q)]$, we may
eliminate the $iz^{0}$ terms. When this is done, the ${\cal O}(\dot{\alpha}^{2})$
term is more straightforward to evaluate; it takes the value
\begin{equation}
\label{eq-dotselfE}
\langle A^{\mu}(z_{2})A^{\nu}(z_{1})\rangle_{\dot{\alpha}^{2}}
=\frac{1}{8}\left(\frac{\dot{\alpha}}{\alpha_{0}}\right)^{2}\int\!\frac{d^{4}q}
{(2\pi)^{4}}e^{-iq\cdot(z_{2}-z_{1})}\frac{-i}{q^{2}}\frac{-i}{q^2}
\left[i\frac{\partial^{2}}{\partial q_{0}^{2}}\Pi_{2}^{\mu\nu}(q)\right].
\end{equation}
The ${\cal O}(\ddot{\alpha})$ term is more complicated. It involves both
derivatives of $\Pi_{2}^{\mu\nu}(q)$ and the external coordinates $z_{1}$ and
$z_{2}$. This kind of dependence of the external coordinates is also present in the
expression for $M(x_{0}+y_{0})$; in that case, the external coordinates just give the
average time between a photon's emission at $z_{1}$ and its absorption at $z_{2}$.
For the second-order terms, it is not so clear what form the
corresponding average should take,
except that at $z_{1}^{0}=z_{2}^{0}$, the self-energy should be
proportional to $\alpha(z^{0})$, the value of the coupling constant at that instant.
The ultimate expression for the ${\cal O}(\ddot{\alpha})$ term is
\begin{eqnarray}
\langle A^{\mu}(z_{2})A^{\nu}(z_{1})\rangle_{\ddot{\alpha}}
& = & \frac{\ddot{\alpha}}{4\alpha_{0}}\left[\left(z_{1}^{0}\right)^{2}
+\left(z_{2}^{0}\right)^{2}\right]\int\!\frac{d^{4}q}{(2\pi)^{4}}
\frac{-i}{q^{2}}\frac{-i}{q^{2}}e^{-iq\cdot(z_{2}-z_{1})}\left[i\Pi_{2}^{\mu\nu}(q)
\right] \nonumber\\
& & +\,\frac{\ddot{\alpha}}{2\alpha_{0}}\int\!\frac{d^{4}q}{(2\pi)^{4}}
\frac{-i}{q^{2}}\frac{-i}{q^{2}}e^{-iq\cdot(z_{2}-z_{1})}\left\{iq_{0}
\frac{\partial}{\partial q_{0}}
\left[\frac{\Pi_{2}^{\mu\nu}(q)}{q^{2}}\right]\right\}.
\label{eq-ddotselfE}
\end{eqnarray}
Together with the lower-order terms, (\ref{eq-dotselfE}) and (\ref{eq-ddotselfE})
give the complete photon propagation amplitude (\ref{eq-completeselfE}).

\end{document}